# Numerical simulation of exciton dynamics in $Cu_2O$ at ultra low temperatures within a potential trap

Sunipa Som, Frank Kieseling and Heinrich Stolz

Universität Rostock, Institut für Physik, Universitätsplatz 3, 18051 Rostock, Germany

E-mail: sunipa.som@uni-rostock.de

**Abstract.** We have studied theoretically the relaxation behaviour of excitons in cuprous oxide ($Cu_2O$) at ultra low temperatures when excitons are confined within a potential trap by solving numerically the Boltzmann equation. As relaxation processes, we have included in this paper deformation potential phonon scattering, radiative and non-radiative decay and Auger decay. The relaxation kinetics has been analysed for temperatures in the range between 0.3K and 5K. Under the action of deformation potential phonon scattering only, we find for temperatures above 0.5K that the excitons reach local equilibrium with the lattice i.e. that the effective local temperature is coming down to bath temperature, while below 0.5K a non-thermal energy distribution remains. Interestingly, for all temperatures the global spatial distribution of excitons does not reach the equilibrium distribution, but stays at a much higher effective temperature. If we include further a finite lifetime of the excitons and the two-particle Auger decay, we find that both the local and the global effective temperature are not coming down to bath temperature.

In the first case we find a Bose-Einstein condensation (BEC) to occur for all temperatures in the investigated range. Comparing our results with the thermal equilibrium case, we find that BEC occurs for a significantly higher number of excitons in the trap. This effect could be related to the higher global temperature, which requires an increased number of excitons within the trap to observe the BEC. In case of Auger decay, we do not find at any temperature a BEC due to the heating of the exciton gas.

# 1. Introduction

The possibility of Bose-Einstein condensation (BEC) of excitons in semiconductors has been an interesting topic for the last few years [1,2]. The exciton, being composed of two fermions, namely an electron and hole, is the basic electronic excited state of an intrinsic semiconductor. Its low effective mass as compared to atoms is a favorable factor for the occurrence of BEC at moderate particle densities.

$Cu_2O$ is a well known semiconductor as a good host for BEC for excitons [3,4]. Due to the special band structure, the exciton in $Cu_2O$ is a ideal model system for kinetic studies. The optical spectrum of $Cu_2O$ is remarkable because all classes of transitions predicted by the theory for exciton spectra are observed in different parts of the energy region in $Cu_2O$. The yellow exciton series is formed from the highest valence band and the lowest conduction band and splits into the triply degenerate orthoexciton and the singly degenerate paraexciton. The energy of paraexciton lies 12 meV lower than the energy of orthoexciton. The high binding energy (150 meV) of excitons in $Cu_2O$ corresponds to a small Bohr radius of $7 \overset{o}{A}$. Due to positive parity of the bands, the orthoexcitons is weakly optically allowed while the paraexciton is optically forbidden. Its decay is only possible via an odd parity optical phonon resulting in a long lifetime in the microsecond range during which thermodynamic-quasi-equilibrium may be reached.

From the previous studies of atomic systems, in which BEC has been demonstrated, we can learn that excitons should be confined in a potential trap. This has the advantage that the diffusion process, which reduces the exciton density is suppressed and the critical number of excitons

$$N_c = \varsigma(3)\left(\frac{k_B T}{\hbar \Omega}\right)^3 \qquad (1)$$

required for the phase transition from exciton gas to BEC, decreases much faster with temperature than in free space [5]. Here $\varsigma$ is the Riemann Zeta function, T the temperature and $\Omega = \sqrt{\frac{a}{M_x}}$ the average oscillator frequency of the trapping potential $V(\vec{r}) = a\vec{r}^2$. In our caclulations we take for the exciton mass $M_x = 2.61 m_e$ [6] and for the steepness constant $a = 0.5 \mu eV \cdot \mu m^{-2}$.

Several groups have reported the observation of Bose distributions of excitons in $Cu_2O$ when using intense photo excitation [7-17]. Thermodynamics of strain-confined paraexcitons in $Cu_2O$ was first studied experimentally in reference [18], while Bosonic stimulation has been observed in [19]. There have also been claims that critical values of density and temperature of a BEC have been obtained either in bulk [12] or in potential traps [14,20]. All these studies, however, did not gave conclusive evidence for a BEC.

The Boltzmann kinetics for both of para and ortho excitons in Cuprous oxide specially for a homogeneous gas have been analyzed numerically and within certain approximations analytically in reference [21,22]. The effects of radiative recombination have been added in reference [22]. The effects of optical-phonon emission, exciton interconversion and Auger decay have been included in reference [23], but it has been assumed that the exciton gas is continuously in internal equilibrium, and followed only the density and temperature of the gas. The relaxation of a spatially homogeneous gas of long-lived excitons under the influence of elastic scattering and LA-phonon emission only has been investigated in reference [24]. Relaxation kinetics of excitons in $Cu_2O$ was studied in reference [25] and [26]. They included all of the known kinetics effects like optical and acoustic phonon emission, Auger decay, elastic scattering and ortho-para-conversion, but not within a potential trap. All of these studies have been undertaken for temperature above 1 K. In contrast, in our work we have studied excitons inside the potential trap and for temperature in the millikelvin range. The kinetics of bosonic excitons in avery weak potential trap was studied in reference [27] by a rate equation analysis in the basis of single-particle eigenfunctions, in contrast in our work in which we solved the Boltzmann equation.

In a typical experiment excitons are created with a kinetic energy on the order of meV and then drift down to the center of the potential trap. Now two questions arise: firstly, how long is the time excitons take to thermalize and secondly, when does a BEC occurs.

We have studied theoretically the relaxation behaviour of excitons in $Cu_2O$ in the temperature range from 0.3 K to 5 K. Recently, this temperature range has been reached by references [20,28]. As we are interested in the behaviour of excitons at long times, it is possible to neglect the orthoexcitons, since they have a rather low concentration at longer times due to their short conversion time into paraexcitons of about 3 ns [25]. We assumed that excitons are confined in a stress induced parabolic potential trap V(r). We did this by solving the Boltzmann equation [29] that describes the statistical distribution of particles in a gas. The Boltzmann equation has been solved by finite difference method and the method of lines [30] using MATLAB.

First we have included deformation potential LA phonon-scattering but no other collisions of excitons. We find that the excitons exhibit a quasi-equilibrium within the potential trap due to the long lifetime on the order of microseconds. An important point is that excitons can thermalize locally within their lifetime for all cases over and below 1 K except at the lowest temperature of 0.3 K. In contrast, globally the excitons don't reach equilibrium with the lattice even at temperatures of 5K, but show always a higher effective global temperature. When we add radiative, non-radiative and Auger decay, we find that both locally and globally the effective temperature is not coming down to bath temperature even after long times.

Our paper is organized as follows: In the section 2 we present details about our numerical modeling. In the section 3 we present our results and compare our results with the results of thermal equilibrium case. The last section is a short summary about our work.

## 2. Numerical Modelling

The Boltzmann equation [28] describes the particle's occupation number $N(\vec{r},\vec{k},t)$ as a function of radius $\vec{r}$, momentum $\hbar\vec{k}$ and time t. The Boltzmann equation is

$$\frac{\partial N}{\partial t} + \vec{v} \cdot \vec{\nabla}_r N + \vec{F} \cdot \frac{1}{\hbar} \vec{\nabla}_k N = \left(\frac{\partial N}{\partial t}\right)_{\text{collision+interaction}} \tag{2}$$

with $\vec{v}$ the velocity, $\vec{F}$ the force and $\vec{\nabla}_r$ and $\vec{\nabla}_k$ the nabla operators in $\vec{r}$ and $\vec{k}$ space, respectively. The terms on the left hand side are often referred to as the drift terms, and the term on the right hand side as collision and interaction term. The collision and interaction terms are phonon scattering, Auger decay, radiative and non-radiative decay and elastic scattering. Here we have included the phonon scattering, Auger decay and radiative and non-radiative decay but elastic scattering is not included in our calculation. For starting the calculation, we have created the excitons, that are actually paraexcitons, which after phonon scattering, Auger decay and radiative and non-radiative decay cool down to the bottom of the trap.

*2.1. Phonon scattering*

First of all we take care of the phonon scattering term, which describes the kinetics of a specially homogeneous system of excitons. We use local approximation and neglect any phonon induced diffusion. The phonon scattering term of the Boltzmann equation [21] is then given by

$$\left(\frac{\partial N_{\vec{k}}}{\partial t}\right)_{\text{phonon-scattering}} = -\frac{2\pi}{\hbar}\sum_{\vec{p}}\left|M_{x-ph}(\vec{p}-\vec{k})\right|^2 \{[N_{\vec{k}}(1+n^{ph}_{\vec{k}-\vec{p}})(1+N_{\vec{p}})$$
$$-(1+N_{\vec{k}})n^{ph}_{\vec{k}-\vec{p}}N_{\vec{p}}]\delta(e_{\vec{k}}-e_{\vec{p}}-\hbar\omega_{\vec{k}-\vec{p}})$$
$$+[N_{\vec{k}}n^{ph}_{\vec{p}-\vec{k}}(1+N_{\vec{p}})-(1+N_{\vec{k}})(1+n^{ph}_{\vec{p}-\vec{k}})N_{\vec{p}}]$$
$$\times\delta(e_{\vec{k}}-e_{\vec{p}}+\hbar\omega_{\vec{p}-\vec{k}})\} . \tag{3}$$

Here $e_{\vec{k}} = \hbar^2 k^2/2M_x$, $e_{\vec{p}} = \hbar^2 p^2/2M_x$ and $\hbar\omega_{\vec{p}-\vec{k}} = \hbar v_s |\vec{p}-\vec{k}|$ are the exciton energy in wavevector $\vec{k}$ state, exciton energy in wavevector $\vec{p}$ state and phonon energy, respectively. $N_{\vec{k}}$, $N_{\vec{p}}$ and $n^{ph}_{\vec{p}-\vec{k}} = 1/[\exp(\hbar\omega_{\vec{p}-\vec{k}}/k_B T_b)-1]$ are the exciton occupation number in $\vec{k}$ state, exciton occupation number in $\vec{p}$ state and phonon occupation number, respectively. $v_s$ is the longitudinal acoustic sound velocity. $M_{x-ph}(\vec{p}-\vec{k})$ is the matrix element of the exciton-phonon deformation potential interaction. The exciton-phonon coupling is given by $\left|M_{x-ph}(\vec{p}-\vec{k})\right|^2 = \hbar D^2 |\vec{p}-\vec{k}|/(2V\rho v_s)$, where D is the deformation potential energy, V is the crystal volume, $\rho$ is the crystal density, $M_x$ is the exciton mass and $\delta$ is the Dirac distribution. The first term in the square brackets on the right-hand side of equation (3) is due to the Stokes scattering and the second term is for anti-Stokes scattering of excitons. The parameters, which we have used, are the effective mass $M_x = 2.61 m_e$, with the electron rest mass $m_e = 9.109 \times 10^{-31}\ kg$, the speed of sound $v_s = 4.5 \times 10^3$ m/s, the deformation potential for LA phonon scattering D=1.68 eV and the crystal density $\rho = 6.11 \times 10^3\ kg/m^3$. Due to symmetry there is no coupling between paraexcitons and TA phonons [25].

From well known results for an isotropic initial distribution of excitons $N_{\vec{k}}$ in momentum space [22], equation (3) can be reduced to a kinetic equation for the exciton distribution function $N_e$ in the one dimensional energy space. For this purpose we transfer all $N_{\vec{k}}$ terms in momentum space to $N_e$ terms in the one dimensional energy space by

$$\frac{\partial N_e}{\partial t} = \frac{V}{(2\pi)^3} \int_0^\pi \sin\theta d\theta \int_0^{2\pi} d\varphi \int_0^\infty k^2 \frac{\partial N_{\vec{k}}}{\partial t} dk \qquad (4)$$

where $k, \theta, \varphi$ are the spherical coordinates of wavevector $\vec{k}$. Inserting the value of $\frac{\partial N_{\vec{k}}}{\partial t}$ in equation (4) and transfer all $\vec{k}$ terms into e terms with $e = \hbar^2 k^2 / 2M_x$ and $e_1 = \hbar^2 p^2 / 2M_x$ we obtain the kinetic energy equation

$$\left(\frac{\partial N_e}{\partial t}\right)_{\text{phonon-scattering}} = -\left(\frac{D^2 \sqrt{M_x}}{4\sqrt{2}\pi\hbar^4 v_s^4 \rho \sqrt{e}}\right) \int_0^\infty de_1 (e-e_1)^2$$
$$\times \{[N_e(1+n^{ph}_{e-e_1})(1+N_{e_1}) - (1+N_e)n^{ph}_{e-e_1} N_{e_1}]\Theta(q_s - \sqrt{e} + \sqrt{e_1})$$
$$\times \Theta(-q_s + \sqrt{e} + \sqrt{e_1})\Theta(e-e_1)$$
$$+ [N_e n^{ph}_{e_1-e}(1+N_{e_1}) - (1+N_e)(1+n^{ph}_{e_1-e})N_{e_1}]$$
$$\times \Theta(q_{as} - \sqrt{e_1} + \sqrt{e})\Theta(-q_{as} + \sqrt{e_1} + \sqrt{e})\Theta(e_1 - e)\} . \qquad (5)$$

$q_{s/as} = \pm(e-e_1)/(\sqrt{2M_x} v_s)$ refers to the momentum transfer of Stokes and anti-Stokes scattering, respectively. The Heaviside step function $\Theta$ in equation (5) is coming from the Dirac distribution in equation (3).

The characteristic scattering time is defined by $\tau_{sc} = (\pi\hbar^4 \rho)/(4D^2 M_x^3 v_s n^{ph}_{e_0})$ as the probability of anti-Stokes scattering of a particle from the ground-state mode, energy equal to zero. Here $e_0 = 2M_x v_s^2$ is the energy of the partner state coupled to the ground-state mode by the emission or absorption of an acoustic phonon. For the ground-state mode, energy equal to zero, equation (3) yields [19]

$$\left(\frac{\partial}{\partial t} N_{e=0}\right)_{\text{phonon-scattering}} = \left(\frac{4D^2 M_x^3 v_s}{\pi\hbar^4 \rho}\right)\left[N_{e_0}(1+n^{ph}_{e_0}) - N_{e=0}(n^{ph}_{e_0} - N_{e_0})\right] . \qquad (6)$$

The first term in the square brackets of equation (6) describes population of the ground-state mode due to phonon assisted spontaneous emission from the partner-state $e_0$, while the second one characterizes the stimulated kinetics proportional to the occupation number $N_{e=0}$.

We discretize the energy $e_j = (j-1)\Delta e$ where $j = 1......ne$ and $\Delta e = \frac{e_0}{M}$ with M is an integer. Analogously we discretize the energy $e_{1q} = (q-1)\Delta e$. In our case $ne=100$. We know that $N^{ph}_{j-q} = 1/[\exp(e_0(j-q)/(k_B TM)) - 1]$, where $T$ is the lattice temperature. So we can write the equation (5) in the form

$$\left(\frac{\partial N_j}{\partial t}\right)_{\text{phonon-scattering}} = -\frac{C}{M^{5/2}\sqrt{j-1}} \sum_{q_{min}}^{q_{max}} (j-q)^2 \times [\{N_j(1+N_{j-q}^{ph})(1+N_q) - (1+N_j)N_{j-q}^{ph}N_q\}$$
$$+ \{N_j N_{q-j}^{ph}(1+N_q) - (1+N_j)(1+N_{q-j}^{ph})N_q\}], \tag{7}$$

with, $C = \dfrac{D^2 M_x^3 v_s}{\pi \hbar^4 \rho}$.

For stokes scattering we obtain $q_{max} = j-1$ and $q_{min} = \left(\sqrt{j-1} - \sqrt{M}\right)^2$ and for anti-stokes scattering $q_{min}$ is the maximum of $q1_{min}$ and $q2_{min}$, where $q1_{min} = j-1$ and $q2_{min} = \left(\sqrt{j-1} - \sqrt{M}\right)^2$ and $q_{max} = \left(\sqrt{j-1} + \sqrt{M}\right)^2$, with $j=2\ldots\ldots ne$. We get these values of $q_{max}$ and $q_{min}$ for stokes and anti-stokes scattering from the Dirac distribution in equation (3).

We can write the equation (6) for $j=1$ in the form

$$\left(\frac{\partial N_{j=1}}{\partial t}\right)_{\text{phonon-scattering}} = \left(\frac{4D^2 M_x^3 v_s}{\pi \hbar^4 \rho}\right) \left[ N_{e_0}\left(1+n_{e_0}^{ph}\right) - N_{j=1}\left(n_{e_0}^{ph} - N_{e_0}\right) \right]. \tag{8}$$

*2.2. Radiative and non-radiative decay*

The radiative and non-radiative decay [28] of paraexcitons occurs with the total rate $\Gamma_p$, i.e.

$$\left(\frac{\partial N_{\vec{k}}}{\partial t}\right)_{\text{R\_NR decay}} = -\Gamma_p N(\vec{r}, e) \tag{9}$$

From recent experiment the decay rate is $\Gamma_p = 1/650$ ns$^{-1}$ [28].

*2.3. Auger decay*

Auger decay [25] destroys two excitons by recombination of one exciton and ionization of the other one. We assume that all of the ionized carriers released by Auger decay rebind to form new excitons, and we distribute these excitons over the whole energy range. Indeed, the electron hole pairs also recombine into orthoexcitons, but these are then converted into paraexcitons. Therefore, only paraexcitons need to be considered. So, in recovery, we are getting back half of the excitons destroyed by Auger decay. The complete effect of Auger decay on the exciton occupation number is [25]

$$\left(\frac{\partial N_{\vec{k}}}{\partial t}\right)_{\text{Auger Decay}} = -A_{pp} n(\vec{r}) N(\vec{r},\vec{k}) + \frac{1}{2} A_{pp} n(\vec{r})^2 \cdot \frac{(2\pi)^3}{\int d^3\vec{k}} \tag{10}$$

or,

$$\left(\frac{\partial N_e}{\partial t}\right)_{\text{Auger Decay}} = -A_{pp}n(\vec{r})N(\vec{r},e) + \frac{1}{2}A_{pp}n(\vec{r})^2 \cdot \frac{(2\pi)^3}{\int_0^{e_{\max}} \frac{4\sqrt{2}\pi M_x^{3/2}\sqrt{e}}{\hbar^3}de}$$

$$= -A_{pp}n(\vec{r})N(\vec{r},e) + \frac{1}{2}A_{pp}n(\vec{r})^2 \cdot \frac{3\hbar^3(2\pi)^3}{8\sqrt{2}\pi M_x^{3/2}e_{\max}^{3/2}} \quad (11)$$

The first term is due to two body decay and the second term is due to recovery. $A_{pp}$ is the Auger constant. $n(\vec{r})$ is the local density obtained by summation of exciton occupation number along e direction.

$$n(\vec{r}) = \frac{1}{(2\pi)^3}\int_k 4\pi k^2 N(\vec{r},\vec{k})dk \quad (12)$$

or,

$$n(\vec{r}) = \frac{1}{(2\pi)^3}\int_e \frac{4\sqrt{2}\pi M_x^{3/2}\sqrt{e}}{\hbar^3} \cdot N(\vec{r},e)de \quad (13)$$

We have used an Auger constant $A_{pp} = 10^{-18}$ cm$^3$/ns taken from recent experiments [28].

*2.4. Drift and force term of the Boltzmann equation*

Drift and force terms of the Boltzmann equation (2) are given by

$$\left(\frac{\partial N}{\partial t}\right)_{\text{Drift \& force}} = -\vec{v}\cdot\vec{\nabla}_r N - \vec{F}\cdot\frac{1}{\hbar}\cdot\vec{\nabla}_k N \quad, \quad (14)$$

where $\vec{v}$ is the velocity and $\vec{F}$ is the force. As $\hbar\vec{k} = M_x\vec{v}$ we have $\vec{v} = \frac{\hbar}{M_x}\vec{k}$.

Further, $\vec{F} = -\vec{\nabla}V(\vec{r})$. With $V(\vec{r}) = a\vec{r}^2$ we have $\vec{F} = -2a\vec{r}$, where $a$ is the steepness constant.

Since the trap and therefore the potential energy has a spherical symmetry it is reasonable if we take a transformation from general coordinates to spherical ones. In this case equation (14) is easily written as

$$\left(\frac{\partial N}{\partial t}\right)_{\text{Drift \& force}} = -\frac{\hbar k}{M_x}\cdot\frac{\partial N}{\partial r} - \frac{-2ar}{\hbar}\cdot\frac{\partial N}{\partial k} \quad. \quad (15)$$

By using finite difference method, we can write

$$\frac{\Delta N}{\Delta r} = \frac{N_i - N_{i-1}}{\Delta r} \text{ , with } r = r_0........r_i \text{ and } r_i = i\Delta r,$$

$$\frac{\Delta N}{\Delta k} = \frac{N_{j+1} - N_j}{\Delta k_j}, \text{ with } k = k_0........k_j \text{ and } k_j = j\Delta k_j, \text{ with } \Delta k_j = \frac{\Delta e\sqrt{M_x}}{\hbar\sqrt{2e_j}}.$$

Using $e = \dfrac{\hbar^2 k^2}{2M_x}$, $e_j = \dfrac{e_0 j}{M}$, $\Delta e = \dfrac{e_0}{M}$ and $e_0 = 2M_x v_s^2$ we get the equation

$$\left(\frac{\partial N}{\partial t}\right)_{\text{Drift \& force}} = -\frac{2v_s}{\sqrt{M}\,\Delta r}\sqrt{j}\{N(i,j) - N(i-1,j)\}$$
$$+\frac{2a\Delta r\sqrt{M}}{M_x v_s} i\sqrt{j}\{N(i,j+1) - N(i,j)\} \quad . \tag{16}$$

*2.5. Complete Boltzmann equation, initial and boundary conditions*

The complete Boltzmann equation is
$$\frac{\partial N}{\partial t} = -\frac{2v_s}{\sqrt{M}\,\Delta r}\sqrt{j}\{N(i,j) - N(i-1,j)\}$$
$$+\frac{2a\Delta r\sqrt{M}}{M_x v_s} i\sqrt{j}\{N(i,j+1) - N(i,j)\} + \left(\frac{\partial N}{\partial t}\right)_{\text{phonon-scattering}}$$
$$+\left(\frac{\partial N}{\partial t}\right)_{\text{R\_NR\_decay}} + \left(\frac{\partial N}{\partial t}\right)_{\text{Auger\_decay}} \tag{17}$$

The boundary conditions are derived from the fact that no exciton flows takes place outside the trap. Therefore
(1) The derivative of exciton occupation number with respect to radius
  (i) $\dfrac{\partial N}{\partial r} = 0$ at $r < 0$, for all values of $e$.
  (ii) $\dfrac{\partial N}{\partial r} = 0$ at $r > r_{max}$, for all values of $e$.
  (iii) $\dfrac{\partial N}{\partial r} = 0$ at $e > e_{max}$, for all values of r.

(2) The derivative of exciton occupation number with respect to energy
  (i) $\dfrac{\partial N}{\partial e} = 0$ at $e < 0$, for all values of $r$.
  (ii) $\dfrac{\partial N}{\partial e} = 0$ at $e > e_{max}$, for all values of $r$.
  (iii) $\dfrac{\partial N}{\partial e} = 0$ at $r > r_{max}$, for all values of $e$.

In our case the maximum radius $r_{max} = 200$ μm and the maximum energy $e_{max} = 6$ meV.

In our calculation, we have used laser excitation which is representative for actual experimental studies [28]. Here the laser beam at energy $e_L$ with respect to the bottom energy of the trap crosses the trap centrally and leads to exciton creation at $e = e_L - a\vec{r}^2$. If $\sigma$ is the spectral width, then the initial exciton occupation number distribution can be formulated as

$$N_0(\vec{r},e) = n_0 \exp\left(\frac{-(e-(e_L - a\vec{r}^2))^2}{2\sigma^2}\right) \Theta(\vec{r}_{max} - \vec{r}) \Theta(\vec{r}) \tag{18}$$

where the constant $n_0$ determines the overall number of excitons and $r_{max} = \sqrt{\frac{2e_L}{a}}$ is the maximum radius. For a temperature of 3 K, we have used $\sigma = 0.12$ meV and the laser energy of $e_L = 2.7$ meV, while for 0.3K and 0.5K $e_L = 0.45$ meV.

## 3. Results

After solving the Boltzmann equation with phonon scattering term we are getting the exciton occupation number as a function of energy $e$, radius $\vec{r}$ and time $t$ (see Figure 1). We see that for a particular temperature, with increasing time, the excitons are going towards the bottom of the trap and after some time they are accumulated in a place near the bottom of the trap. In this case the total exciton number within the trap is $1.7 \times 10^4$.

We have calculated the total number of excitons by the equation

$$N_{total} = \frac{1}{(2\pi)^3} \int_0^\infty \int_0^\infty 4\pi r^2 dr \cdot 4\pi k^2 dk \cdot N(\vec{r},\vec{k}) \; . \tag{19}$$

Further quantities are the summed number of excitons along $r$ direction which is given by

$$n_r(e) = \int_0^\infty 4\pi r^2 dr \cdot N_e(\vec{r},e) \tag{20}$$

and the summed number of excitons along the direction of energy given by

$$n_e(\vec{r}) = \frac{1}{(2\pi)^3} \int_0^\infty 4\pi k^2 dk \cdot N(\vec{r},\vec{k}) \quad . \tag{21}$$

While $n_e(\vec{r})$ gives the spatial density distribution of excitons, $n_r(e)$ represents the energy distribution averaged over the trap. From $n_r(e)$ we get the total number of exciton by the equation

$$N_{total} = \frac{1}{(2\pi)^3} \cdot \left(\frac{2M_x}{\hbar^2}\right)^{3/2} \int_e \sqrt{e} \cdot de \cdot n_r(e) \tag{22}$$

In a real experiment [28] the photoluminescence intensity which is proportional to the number of excitons, is measured as a function of location z and the spectral position $\hbar\omega = e + e_0$, where $e_0$ is the minimum energy of the trap. From the Boltzmann equation we get the exciton distribution $N(\vec{r},e')$ as a function of local energy $e' = \hbar\omega - a\vec{r}^2 - e_0$ and radius $\vec{r}$. Then we transfer $\vec{r}$ to the Cartesian coordinates x, y and z and from the distribution $N(\vec{r},e')$ we have calculated the distribution $N(z,e')$, where $N(z,e')$ is exciton occupation number as a function of energy $e'$ and location z. We see only the middle of the trap, therefore in our case x is equal to zero and the distance, that is actually the projected slit width, is $\Delta x$. Then by integrating $N(z,e')$ over the location y we get the energy distribution

$$\frac{N(z,e)}{\Delta x} = \int_{-\infty}^{\infty} 4\pi k^2 \cdot \frac{dk}{de} \cdot N(r = \sqrt{y^2 + z^2}, e')dy \qquad (23)$$

or,

$$\frac{N(z,e)}{\Delta x} = 2\int_{0}^{\infty} \frac{4\sqrt{2}\pi M_x^{3/2}}{\hbar^3} \cdot \sqrt{e} \cdot N\left(r = \sqrt{y^2 + z^2}, e'\right)dy \quad . \qquad (24)$$

Then we have integrated $\frac{N(z,e)}{\Delta x}$ over the z direction and got the energy distribution

$$\frac{N(e)}{\Delta x} = \int_{-\infty}^{\infty} N(z,e)dz = 2\int_{0}^{\infty} N(z,e)dz \qquad (25)$$

which is proportional to the experimentally measured spectrum.

*3.1 Acoustic phonon scattering*

*3.11 Low number of excitons*

In figure 2 we have plotted the final distribution $n_r$ at 100 ns by summation of the total exciton number of non degenerate excitons in r direction vs. energy for different bath temperatures. While for temperatures between 0.5 K and 3 K we obtain a straight line corresponding to thermal equilibrium, for the case of 0.3 K even at 600 ns, which would correspond to the experimental lifetime of reference [28], we do not obtain straight curve. Therefore, in this case we do not get thermal equilibrium.

The inset of figure 2 shows the total number of excitons $N_{tot}$ vs. time for different bath temperatures including phonon scattering. We see that in the beginning for all temperatures the exciton number decreases a little with time and then it stays almost constant. As origin of this initial decay we found that the excitons are scattered out of the finite energy space due to the somewhat singular initial distribution. Reference [31] reported about the same problem and therefore renormalized the exciton number at each time step. To get rid of this numerical artifact, we always considered in our numerical simulations the results for the actual number of excitons at each time. Therefore, our conclusions do not depend on the actual initial number of excitons.

Figure 3 shows the spatial distribution $n_e$ of excitons for 0.5 K as a function of time, i.e. the summation of the exciton number over the direction of energy vs. radius r, with the phonon scattering only. Full lines are the results of the numerical simulations and dash dotted curves are for the thermal equilibrium case at each time. Here we see the deviation between our numerical simulation results and the thermal equilibrium case. From the full line curves we have calculated the global effective temperatures by using the relation of the exciton distribution $n_e(\vec{r}) \sim e^{-\left(\frac{r}{\xi}\right)^2}$ with $\xi = \sqrt{\frac{k_B T_{global}}{a}}$ and the steepness constant $a = 0.5 \mu eV \cdot \mu m^{-2}$. Therefore, we take the value of $\vec{r}$ at the half maximum of the exciton distribution, and determined the effective temperature. In this case the global effective temperature is not coming down to bath temperature.

Then we have calculated the local effective temperature $T_{local}$ for a non degenerate case by fitting the long energy tail of the exciton distribution by the function proportional to $e^{-e/k_B T_{local}}$ and studied how $T_{local}$ changes with time. We have calculated the summation of excitons number $n_r$ along the r direction, fit it with the function that is proportional to $e^{-e/k_B T_{local}}$ and get the values of $T_{local}$ for different bath temperatures (see Figure 4). We see that for temperatures between 1 K and 5 K the effective temperature is coming down to bath temperature within ten nanoseconds. This is different for temperatures below 1K, where the effective temperature is coming down to bath temperature very slowly in around hundreds of nanoseconds only.

The upper inset of figure 4 shows the $T_{homo}$ vs. time for different bath temperatures in homogeneous case. We see the cooling time is same with and without the potential trap.

The lower inset shows the $T_{global}$ vs. time for different bath temperatures within the trap. We see that at longer time for the bath temperature of 0.5 K, $T_{global}$ is coming down to 0.8 K, and for the bath temperature 3 K, $T_{global}$ is coming down to 4 K.

*3.12 High number of excitons and possibility of BEC*

At a sufficiently high exciton number in the trap we expect a BEC to occur. For this purpose we did a number of subsequent simulations with increasing constant $n_0$ in the initial distribution. As an indicator for a BEC we consider a high peak at or near zero energy, i.e. a high number of excitons at the minimum of the potential trap.

A typical set of results is shown in figure 5. Here we have plotted $\frac{N(e)}{\Delta x}$ vs. energy curves with the temperature of 0.5 K with different exciton numbers within the trap and for different times. We see that if the initial exciton number within the trap is $1.02 \times 10^9$ and at 10 ns exciton number is $1.86 \times 10^8$ then no BEC occurs (see figure 5a). In the next two steps we increase the exciton number by the total factor of 8. If the initial exciton number within the trap is $4.08 \times 10^9$ and at 10 ns exciton number is $3.05 \times 10^8$, still no BEC occurs (see figure 5b), but finally if the initial exciton number within the trap is $8.16 \times 10^9$ and at 10 ns exciton number is $4.61 \times 10^8$, then at 10 ns the BEC occurs (see figure 5c) and it is stable over a long time. As we expect that the critical number for BEC depends on the time intervall that the excitons relax in the trap, we have done a large range of simulations with different initial exciton numbers. The results are shown in Figure 6, where the total number of excitons $N_{total}$ within the trap for which the BEC appears first is plotted vs. time for bath temperatures of 0.5 K, 1 K, 3 K and 5 K. Here we see indeed a reduction of the exciton number with time by more than one order of magnitude. In the figure we compare the critical number for BEC from the simulations $N_{total}$ with that expected in thermal equilibrium $N_c$. This is given by

$$N_c = 1.202 \left( \frac{k_B T}{\hbar \Omega} \right)^3 \tag{26}$$

with the frequency $\Omega = \sqrt{\frac{a}{M_x}}$, the exciton mass $M_x = 2.61 m_e$, the steepness constant $a = 0.5 \mu eV \cdot \mu m^{-2}$ and the temperature T.

While we see that from our numerical simulations BEC to occur for all temperatures in the investigated range of 0.5 K to 5 K, the critical numbers are more than an order of magnitude higher than that obtained by assuming the excitons to be in thermal equilibrium with the lattice at T=T$_{Bath}$ (see blue full lines in Figure 6). If we however take into account that the spatial distribution of excitons correspond to a much higher global temperature T$_{global}$ and calculate the corresponding critical numbers N$_c$(T$_{global}$), we obtain a much better agreement with the simulations (see the black dashed lines in Figure 6). From this we can conclude that in order to obtain experimentally a BEC it will be important to cool down the excitons both locally and globally to the lattice temperature.

*3.2 Auger decay*

Then we add radiative and non-radiative decay and Auger decay into the Boltzmann equation. Several groups reported the values of Auger constant. The reported values are, a temperature independent Auger constant of $7 \times 10^{-17}$ cm$^3$ns$^{-1}$ [8, 32], a temperature dependent Auger constant of $1.8 \times 10^{-16}$ cm$^3$ns$^{-1}$ at 77 K [33], a temperature independent Auger constant of $4 \times 10^{-16}$ cm$^3$ns$^{-1}$ in bulk crystals [34]. Furthermore, the experimental results differ by at least four orders of magnitude from the theoretical predictions [33, 35]. Recently, it was reported that Auger decay is actually related to the formation of a biexciton state [36]. As our aim is to model the experimental work of Ref. [28], we take the Auger constant from this report as t $A_{pp} = 10^{-18}$ cm$^3$/ns.

After solving the Boltzmann equation with phonon scattering, radiative and non-radiative decay and Auger decay with the Auger constant $A_{pp}$, we see, that locally the effective temperature is not coming down to the bath temperature even after 600 ns (see figure 7). We have also calculated the global effective temperature (see the inset of figure 7) from the density n$_e$(r) vs. radius curves by the same way like before. Globally, the effective temperature is also not coming down to the bath temperature. In figure 8 we compare the total number of excitons within the trap vs. energy for the bath temperature of 3 K with and without Auger decay and radiative and non-radiative decay with the same initial exciton numbers within the trap. Here we see that without Auger decay and radiative and non-radiative decay the BEC exist at 15 ns (see figure 8a) but with Auger decay and radiative and non-radiative decay no BEC occurs at any time (see figure 8b). A similar behaviour was found for other temperatures.

**4. Conclusions**

We have studied the relaxation behaviour of excitons between 0.3 K to 5 K in a potential trap. By including only the phonon scattering, we see, that for the bath temperatures between 1 K to 5 K the local effective temperature T$_{eff}$ is coming down to bath temperature very quickly within 10 nanoseconds, but for bath temperatures below 1 K, T$_{eff}$ is coming down to bath temperature very slowly within several 100 nanoseconds. This effect can be related to the freezing out of phonons for very low temperatures. On a global scale, the effective exaction temperature T$_{global}$ is not coming down to the bath temperature for all cases considered, but the radial distribution is still a Gaussian. For low bath temperatures the global effective temperature is almost a factor of 2 larger than the bath temperature.

We see from our numerical simulations, that the BEC occurs for all observed temperatures. By comparing our results with the thermal equilibrium case, we see that for all temperatures the BEC comes at a much higher number of excitons than in thermal equilibrium case. However, the effective temperature at the time of BEC is in good agreement with the global effective temperature.

When we add radiative, non-radiative and Auger decay to phonon scattering, we see locally and globally that for all cases the effective temperature is not coming down to the bath temperature even after long times due to the heating of the exciton gas. In this case, no BEC occurs. Therefore the mechanism of relaxation explosion of [20] seems highly implausible.

In this work we have included phonon scattering, radiative and non-radiative decay and Auger decay on the dynamics of excitons. In our future work, it will be important to include elastic scattering of excitons in order to obtain a complete picture of the relaxation dynamics.

**Table 1.** Comparison between $N_c(T_{bath})$, $N_c(T_{global})$ and $N_{total}$.

| $T_{bath}$ (K) | $T_{global}$ (K) | $N_c(T_{bath})$ | $N_c(T_{global})$ | $N_{total}$ |
|---|---|---|---|---|
| 0.5 | 0.9 | $5.45 \times 10^7$ | $3.16 \times 10^8$ | $4.61 \times 10^8$ |
| 1 | 1.4 | $4.36 \times 10^8$ | $1.2 \times 10^9$ | $1.97 \times 10^9$ |
| 3 | 4.6 | $1.18 \times 10^{10}$ | $4.21 \times 10^{10}$ | $6.5 \times 10^{10}$ |
| 5 | 5.43 | $5.45 \times 10^{10}$ | $6.93 \times 10^{10}$ | $9.34 \times 10^{10}$ |

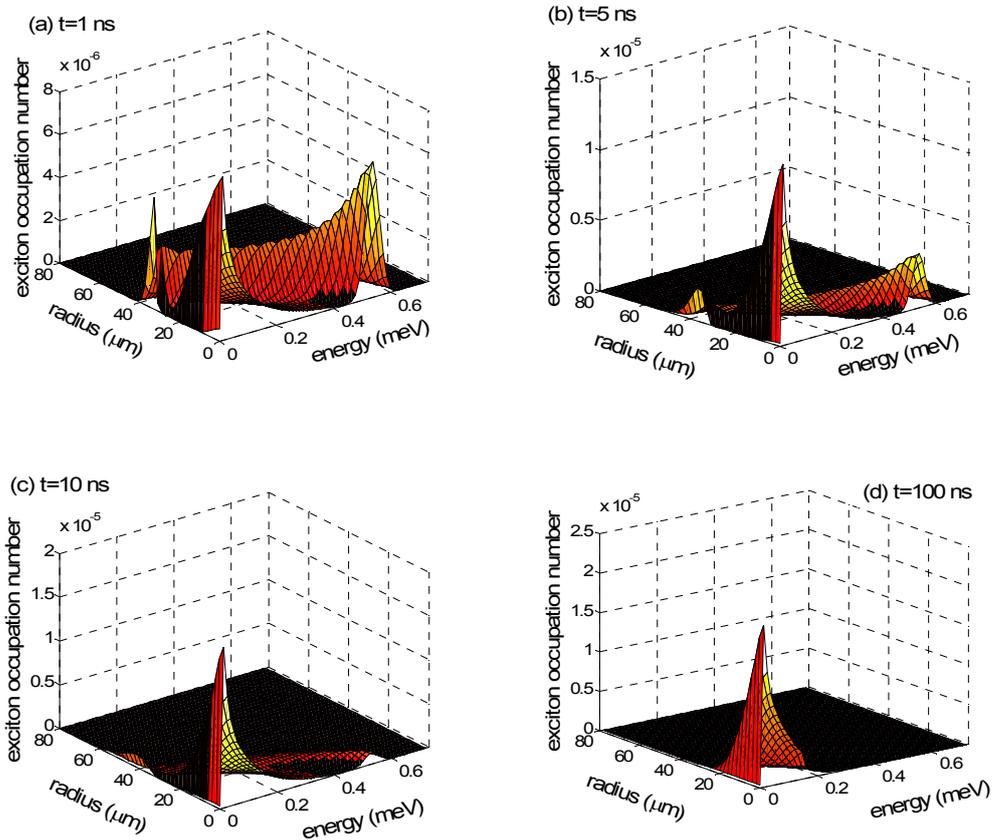

**Figure 1.** The results of the simulation with phonon scattering at different times for 0.5 K when the initial exciton number within the trap is $1.7 \times 10^4$. The series of figures shows that excitons are accumulated near the bottom of the trap within 100 ns at 0.5 K.

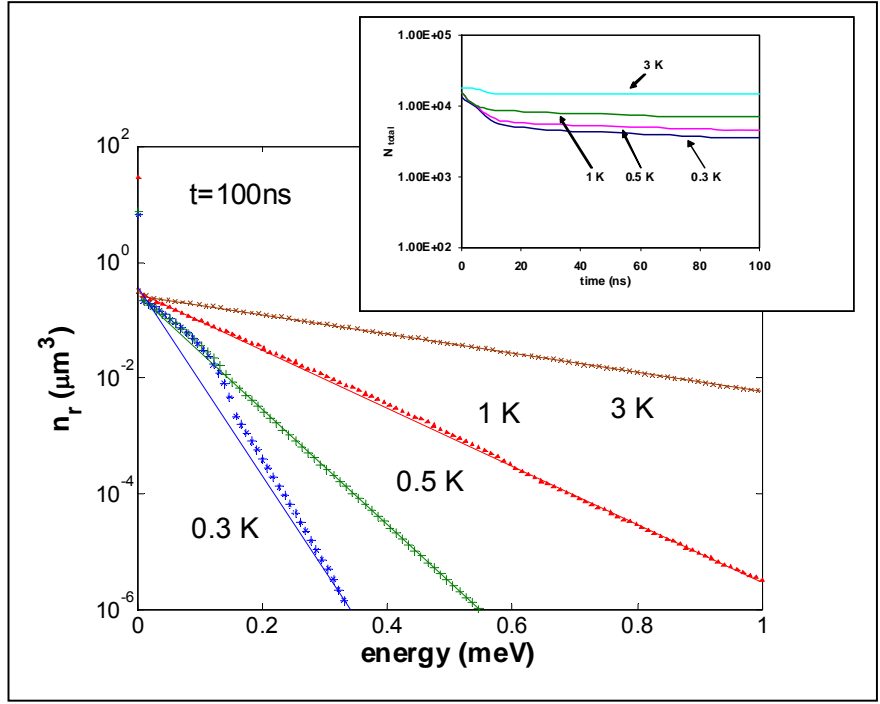

**Figure 2.** The final distribution $n_r$ at 100 ns by summation of exciton number of non-degenerate exciton gas in $\vec{r}$ direction vs. energy for different temperatures. In the case of 0.3 K the exciton distribution is shown at 600 ns. For 0.3 K, 0.5 K, 1 K and 3 K, initial total number of excitons are $1.3 \times 10^4$, $1.5 \times 10^4$, $1.5 \times 10^4$ and $1.8 \times 10^4$, respectively. The markers represent the result of numerical simulations and the full lines represent $n_r$ vs. energy for the thermal equilibrium case. The inset shows the total number of excitons vs. time for different bath temperatures including deformation potential phonon scattering only. We see that in the beginning for all temperatures the exciton number decreases very little with time and then it stays almost constant. As origin of this initial decay we found that the excitons are scattered of the finite energy space due to the somewhat singular initial distribution. For this reason we consider this as numerical artifact.

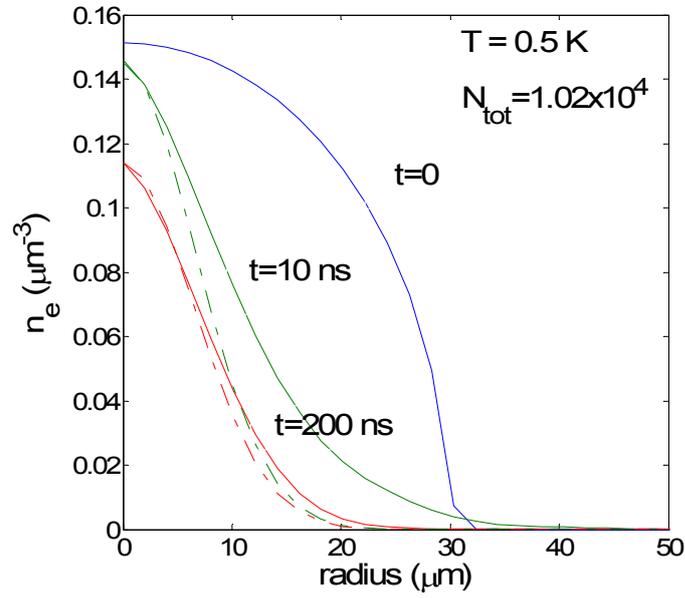

**Figure 3.** Spatial distribution $n_e$ of excitons for 0.5 K as a function of time for non degenerate case, i.e. the summation of exciton number over the direction of energy vs. radius r, with phonon scattering only. For 0.5 K, initial total number of excitons are $1.02 \times 10^4$. Full lines represent the results of numerical simulation and dash-dotted curves represent $n_e$ vs. r for the thermal equilibrium case.

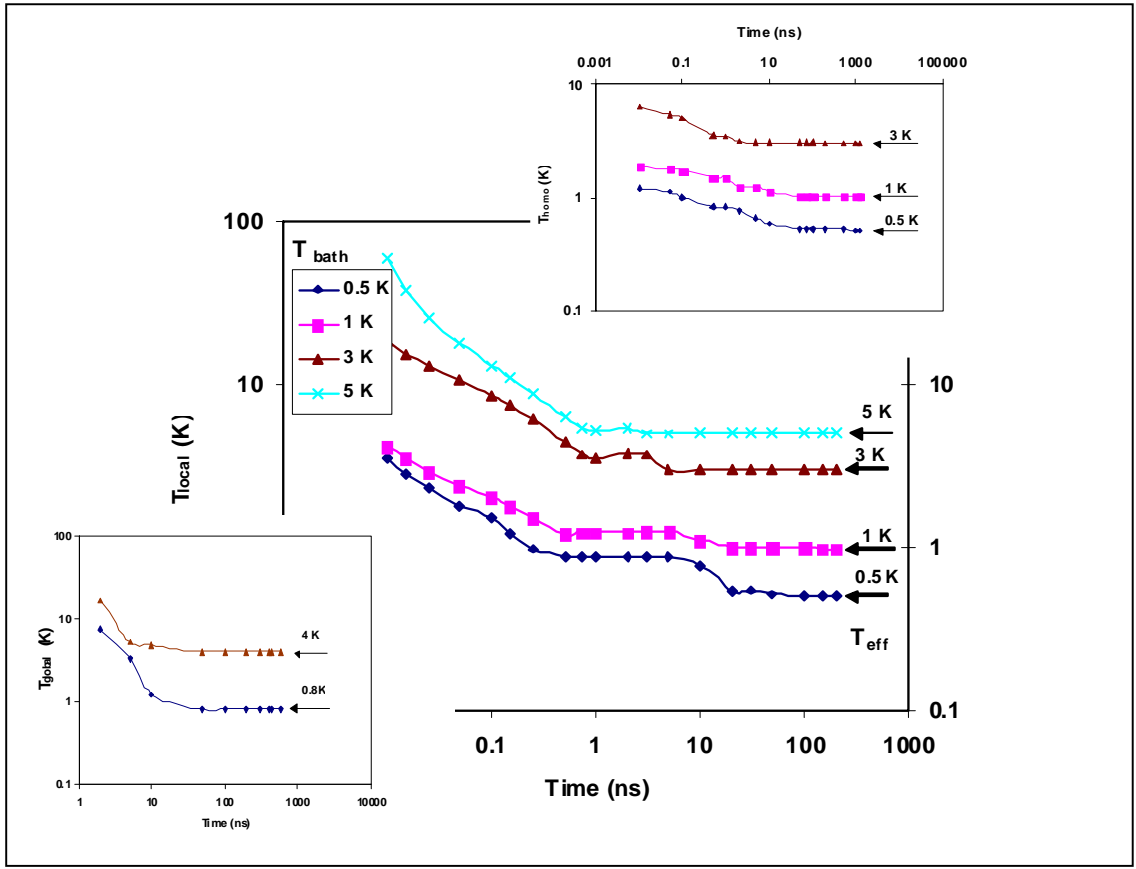

**Figure 4.** Local exciton temperature vs. time for different bath temperatures including deformation potential phonon scattering only. The arrows on the right hand side indicate the effective temperatures. For 0.5 K, 1 K, 3 K and 5 K initial total number of excitons are $1.5 \times 10^4$, $1.5 \times 10^4$, $1.98 \times 10^6$ and $1.99 \times 10^6$, respectively. Here we see that the local effective temperatures are coming down to the bath temperatures. The upper inset shows the $T_{homo}$ vs. time for different bath temperatures in homogeneous case where the local effective temperatures are also coming down to the bath temperatures. The lower inset shows the $T_{global}$ vs. time for different bath temperatures within the trap where the global effective temperatures are not coming down to the bath temperatures. For the calculation of local and global effective temperatures see the text. The curves serve as a guide to the eye.

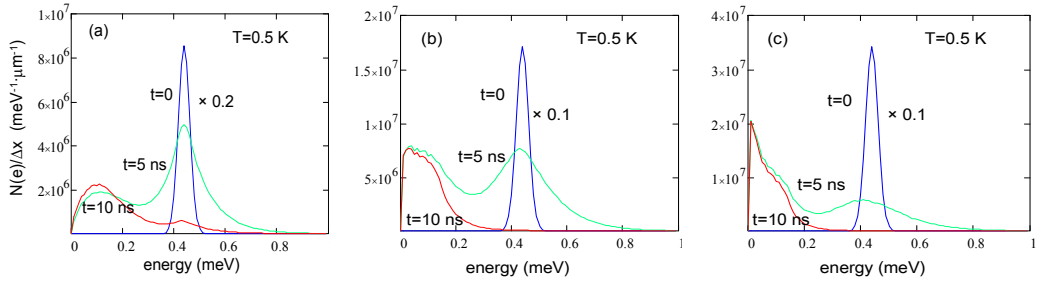

**Figure 5.** The spectral distribution $N(e)/\Delta x$ of excitons as a function of time vs. energy at 0.5 K for a non degenerate case. Figure 5a represents number of excitons within the trap vs. energy for 0.5 K at different times when initial exciton number within the trap is $1.02 \times 10^9$, at 10 ns exciton number is $1.86 \times 10^8$ and $\Delta x = 1\,\mu$m. Figure 5b represents $N(e)/\Delta x$ vs. energy spectrum for 0.5 K at different times when initial exciton number within the trap is $4.08 \times 10^9$ and at 10 ns exciton number is $3.05 \times 10^8$. In both cases we can not see any BEC. Figure 5c represents $N(e)/\Delta x$ vs. energy for 0.5 K at different times when initial exciton number within the trap is $8.16 \times 10^9$ at 10 ns exciton number is $4.61 \times 10^8$. In figure 5c at 10 ns a high peak near zero energy indicates the occurence of BEC.

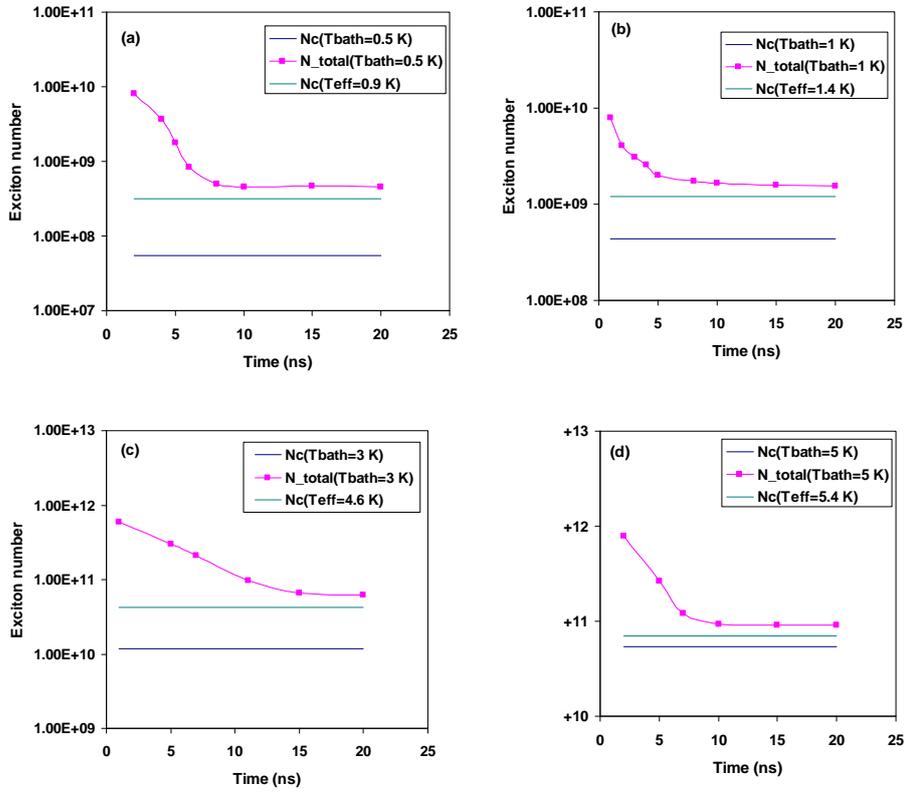

**Figure 6.** Exciton number vs. time curves for different temperatures. Red curves represent total number of excitons within the trap for which for the first time BEC appears vs. time for 0.5 K, 1 K, 3 K and 5 K. In these cases exciton number is decreasing with time and coming to the BEC. Green curves represent critical number of excitons at global effective temperature for thermal equilibrium case. Blue curves represent the critical number of excitons at bath temperature for thermal equilibrium case.

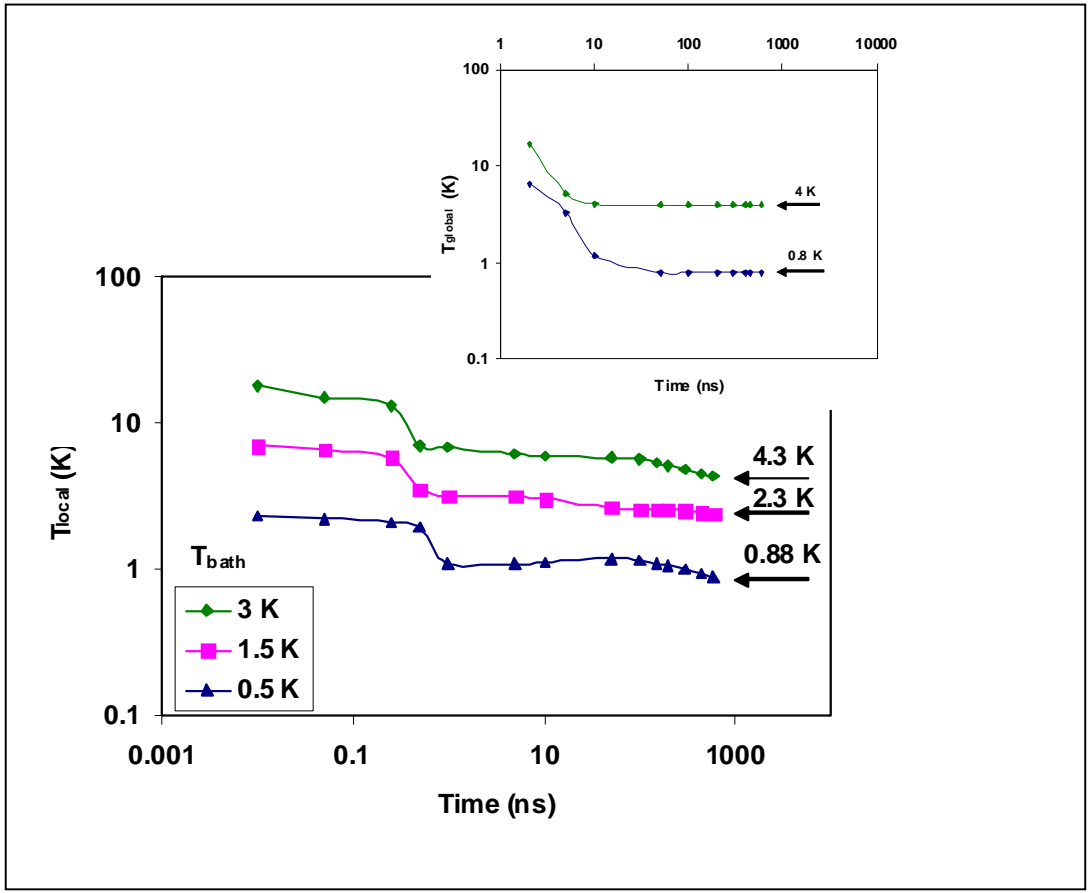

**Figure 7.** Local exciton temperature vs. time curves for different bath temperatures including deformation potential phonon scattering, radiative and non-radiative decay and Auger decay. The arrows on the right hand side indicate the effective temperatures. We see that the effective temperatures are not coming down to bath temperatures even after 600 ns. At 0.5 K, 1.5 K and 3 K initial exciton number within the trap are $1.02 \times 10^9$, $2.76 \times 10^{10}$ and $1.99 \times 10^{11}$, respectively. The inset shows $T_{global}$ vs. time curves for different bath temperatures where we see that the global effective temperatures are not also coming down to the bath temperatures. For the calculation of local and global effective temperatures see the text. The curves serve as a guide to the eye.

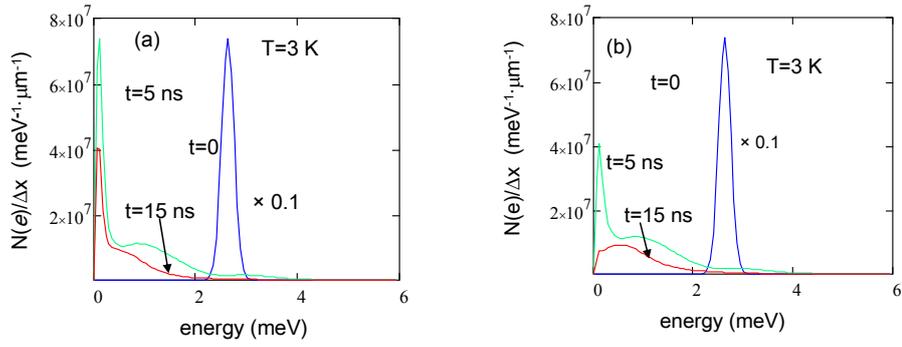

**Figure 8.** The spectral distribution N(*e*)/ Δx of excitons as a function of time vs. energy at 3 K with and without Auger decay and radiative and non-radiative decay. Figure 8a represents the results only with phonon scattering. Here initial excitons number within the trap is $5.97 \times 10^{11}$ and at 15 ns exciton number is $6.5 \times 10^{10}$. At 15 ns, we see that a high peak near zero energy occurs which indicates BEC. Figure 8b represents the results including Auger decay and radiative and non-radiative decay with phonon scattering, and the same exciton number within the trap. In this case we don't see any BEC at any time.